\documentclass[11pt]{article}

\usepackage{amsmath}
\usepackage{fullpage}
\usepackage{amssymb}
\usepackage{amsthm}
\usepackage{hyperref}

\newcommand{\comment}[1]{}
\newcommand{\ket}[1]{|#1\rangle}
\newcommand{\bra}[1]{\langle #1|}
\newcommand{\braket}[2]{\langle #1 | #2 \rangle}

\newcommand{\norm}[1]{\left\|#1\right\|}

\newcommand{\eps}{\epsilon}
\renewcommand{\Pr}{\mathbb{P}}
\newcommand{\Expect}{\mathbb{E}}
\newcommand{\tr}{\operatorname{tr}}

\newcommand{\ot}{\otimes}
\newcommand{\cE}{\mathcal{E}}

\newcommand{\cF}{\mathcal{F}}
\newcommand{\cU}{\mathcal{U}}
\newcommand{\cP}{\mathcal{P}}
\newcommand{\cN}{\mathcal{N}}
\newcommand{\cH}{\mathcal{H}}

\newcommand{\be}{\begin{equation}}
\newcommand{\ee}{\end{equation}}
\newcommand{\bes}{\begin{equation*}}
\newcommand{\ees}{\end{equation*}}
\def\ba#1\ea{\begin{align}#1\end{align}}
\def\bas#1\eas{\begin{align*}#1\end{align*}}
\def\bit{\begin{itemize}}
\def\eit{\end{itemize}}
\def\l{\left}
\def\r{\right}
\def\<{\langle}
\def\>{\rangle}

\newtheorem{theorem}{Theorem}
\newtheorem{lemma}[theorem]{Lemma}
\newtheorem{definition}[theorem]{Definition}
\newtheorem{corollary}[theorem]{Corollary}

\numberwithin{equation}{section}
\numberwithin{theorem}{section}

\newcommand{\eq}[1]{Eqn.~\ref{eq:#1}}

\newcommand{\thmref}[1]{Theorem \ref{thm:#1}}

\newcommand{\lemref}[1]{Lemma \ref{lem:#1}}
\newcommand{\secref}[1]{Section \ref{sec:#1}}
\newcommand{\defref}[1]{Definition \ref{def:#1}}

\DeclareMathOperator{\poly}{poly}

\def\bbE{\mathbb{E}}
\def\Pr{\mathbb{P}}

\begin{document}

\title{Large Deviation Bounds for $k$-designs}

\author{Richard A.~Low\footnote{low@cs.bris.ac.uk}\\Department of Computer Science, University of Bristol, Bristol, U.K.}
\maketitle

\begin{abstract}

We present a technique for derandomising large deviation bounds of functions on the unitary group.  We replace the Haar distribution with a pseudo-random distribution, a $k$-design.  $k$-designs have the first $k$ moments equal to those of the Haar distribution.  The advantage of this is that (approximate) $k$-designs can be implemented efficiently, whereas Haar random unitaries cannot.  We find large deviation bounds for unitaries chosen from a $k$-design and then illustrate this general technique with three applications.  We first show that the von Neumann entropy of a pseudo-random state is almost maximal.  Then we show that, if the dynamics of the universe produces a $k$-design, then suitably sized subsystems will be in the canonical state, as predicted by statistical mechanics.  Finally we show that pseudo-random states are useless for measurement based quantum computation.

\end{abstract}

\section{Introduction}

There are many results in quantum information theory that show generic properties of states or unitaries (e.g.~\cite{AspectsOfGenericEntanglement, RandomizingQuantumStates04}).  Often, these results say that, with high probability, a random state or unitary has some property, for example high entropy.  However, simple parameter counting shows that random unitaries cannot be obtained efficiently.  This limits the usefulness of such results since no physical systems will behave truly randomly.  To make such results more physically relevant, it would be desirable to show that these properties are generic properties of unitaries from some natural distribution that can be implemented efficiently.  Only then could we conclude that we would expect to see such properties in natural systems.

In many cases, the generic properties of unitaries are desirable but randomised constructions given by the large deviation bounds are inefficient.  We would like to come up with distributions which can be implemented efficiently that have similar generic properties.  Two examples where the best known constructions are inefficient randomised ones are $\infty$-norm randomising maps \cite{RandomizingQuantumStates04, AubrunRandomizingChannels} and locking of classical correlations \cite{Locking04, RandomizingQuantumStates04}.  The results show that, with some non-zero probability, random unitaries have the required property.  However, there are no known efficient constructions of unitaries with these properties.  If, on the other hand, we could show that unitaries drawn randomly from a set that can be implemented efficiently have the property with non-zero probability, we could move an important step closer to finding efficient constructions.  (It would not actually provide an efficient construction unless we could find an efficient sampling method.)  In fact, for the case of $\infty$-norm randomisation, this was done by Aubrun in \cite{AubrunRandomizingChannels}.  

By random unitaries, we mean unitary matrices distributed according to the unitarily invariant Haar measure.  In this paper, we will consider replacing the Haar measure with a $k$-design.  A $k$-design is an ensemble of unitaries such that the $k^{\text{\rm th}}$ moments are the same as for the Haar measure \cite{DCEL06}.  ($k$-designs are formally defined in \secref{kDesigns}).  In particular, this means that the expectation of a polynomial in the elements of the unitary matrices of degree at most $k$ is the same whether the distribution is the Haar measure or a $k$-design.

The reason for using $k$-designs is two-fold.  Firstly, because the first $k$ moments are the same we would expect similar (although weaker) measure concentration results.  Secondly, for $k = \poly(n)$ (when the design is on $n$ qubits), we might expect to be able to implement the $k$-design efficiently (i.e.~in $\poly(n)$ time).  Indeed, for $k = O(n/\log n)$, \cite{EfficientTPE08} provides an efficient $k$-design construction\footnote{Only when we allow for approximate designs, however, we can make the approximation good enough to not significantly affect the results.}. 

Not only can $k$-designs be constructed efficiently, they may even be the product of generic dynamics.  In \cite{RandomCircuits08}, it is shown that random quantum circuits quickly converge to a 2-design for a quite general model of such circuits.  It is also conjectured in \cite{RandomCircuits08} that random circuits give $k$-designs for $k>2$ and $k=\poly(n)$ in polynomial time.  If a physical system can be accurately modelled by a random circuit then, assuming this conjecture, the naturally occurring states will be $k$-designs rather than fully random states.

We now summarise some related results in this area.  Smith and Leung \cite{SmithLeung06} and Dahlsten and Plenio \cite{DahlstenPlenio05} found large deviation bounds for stabiliser states.  They showed that, in certain regimes, stabiliser states are very likely to have large entanglement.  Stabiliser states are 2-designs so our results can be seen as a generalisation of this to $k$-designs for $k>2$ and to other problems.  There are also some recent classical results related to the present work.  Alon and Nussboim \cite{AlonNussboim08} consider replacing full randomness with $k$-wise independence, a classical analogue of $k$-designs, in random graph theory.  They show that $k$-wise independent random graphs with $k = \poly(\log N)$ ($N$ is the number of vertices) have similar generic properties to fully random graphs.

\subsection{Introductory Problem: Entanglement of a 2-design}
\label{sec:EntanglementOfA2design}

We now illustrate our main idea by showing a large deviation bound for the entanglement of a 2-design, but in a different way to \cite{SmithLeung06, DahlstenPlenio05}.

It has been known for a long time that random states are highly entangled across any bipartition \cite{PagesConjecture, PagesConjectureProof94, PagesConjectureProof95}.  Further, in \cite{AspectsOfGenericEntanglement}, it is shown that random unitaries generate almost maximally entangled states with high probability.  However, generating random states is inefficient so it is an interesting question to ask if random efficiently obtainable states are highly entangled.

Let the system be $\cH = \cH_S \ot \cH_E$, where we label the two systems $S$ and $E$.  Let the dimensions be $d_S$ and $d_E$ and $d = d_S d_E$.  Let the overall initial state be any fixed state $\rho_0$.  Then consider applying a random unitary $U$ to $SE$ to get the state $\psi = U \rho_0 U^\dagger$.  Then the von Neumann entropy $S(\psi_S)$ of the reduced state $\psi_S = \tr_E \psi$ is close to $\log_2 d_S$ (the maximal) with high probability:
\begin{theorem}[\cite{AspectsOfGenericEntanglement} Theorem 3.3]
\label{thm:EntropyTailBoundFullRandomness}
Let $d_E \ge d_S \ge 3$.  Then for unitaries chosen from the Haar measure
\be
\Pr(S(\psi_S) \le \log_2 d_S -\alpha -\beta) \le \exp \l(- \frac{ (d-1)C \alpha^2}{(\log_2 d_S)^2} \r)
\ee
where $C = \frac{1}{8 \pi^2}$ and $\beta = \frac{1}{\ln 2} \frac{d_S}{d_E}$.
\end{theorem}
Now, consider choosing the unitary from a 2-design instead.  Later on (\lemref{ExpectedPurity}), we show that $\bbE \tr \psi_S^2 = \frac{d_S + d_E}{d+1} =: \mu$.  Since purity is a polynomial of degree $2$, it does not matter if we take the expectation over the Haar measure or the 2-design.  We now apply Markov's inequality:
\bas
\Pr\l(\tr \psi_S^2 \ge \mu\gamma\r) &\le \frac{\bbE \tr \psi_S^2}{\mu\gamma} \\
&= \frac{1}{\gamma}.
\eas
Using the bound $S(\psi_S) \ge -\log_2 \tr \psi_S^2$ and some manipulations (the details are in \secref{Entropy}), this can be written as
\be
\Pr(S(\psi_S) \le \log_2 d_S -\alpha -\beta) \le 2^{-\alpha}
\ee
where $\beta$ is as in \thmref{EntropyTailBoundFullRandomness}.  This bound is much weaker than the bound in \thmref{EntropyTailBoundFullRandomness} and, in particular, does not show strong concentration as $d$ increases.  Later in the paper, we will show that choosing unitaries from a $k$-design with larger $k$ will give a much stronger bound that does give sharp concentration results for large $d$.

\subsection{Main Results}

We will now state our main results.  In the remainder of the paper we will use the following notation to identify the distribution we are using.  When the unitaries are chosen from a distribution $\nu$, we will write $\Pr_\nu$ to mean $\Pr_{U \sim \nu}$, the probability when $U$ is chosen from $\nu$.  Similarly for $\bbE_\nu$, the expectation.  Usually $\nu$ will be a $k$-design.  When the distribution is the Haar measure, we will write a subscript $H$.  So for the Haar average we write $\bbE_{H}$ for $\bbE_{U \sim \cU(d)}$.

Our most general result is:
\begin{theorem}
\label{thm:ConcentrationPolynomial}
Let $f$ be a polynomial of degree $K$.  Let $f(U) = \sum_i \alpha_i M_i(U)$ where $M_i(U)$ are monomials and let $\alpha(f) = \sum_i | \alpha_i |$.  Suppose that $f$ has probability concentration
\begin{equation}
\label{eq:concentration}
\Pr_H(|f - \mu| \ge \delta) \le C e^{-a \delta^2}
\end{equation}
and let $\nu$ be an $\eps$-approximate unitary $k$-design.
Then
\begin{equation}
\Pr_\nu(|f - \mu| \ge \delta) \le \frac{1}{\delta^{2m}} \left( C \left(\frac{m}{a}\right)^m + \frac{\eps}{d^k} \left(\alpha + | \mu | \right)^{2m} \right)
\end{equation}
for integer $m$ with $2mK \le k$.
\end{theorem}
We therefore take a bound for Haar random unitaries of the form \eq{concentration} and turn it into a bound for $k$-designs.  For our definition of $\eps$-approximate designs, see \secref{kDesigns}.  Often, we will use Levy's Lemma (\lemref{Levy}) to give the initial concentration bound in \eq{concentration}.  In this case, $a = \Theta(d)$ (provided the Lipschitz constant (see later) is constant).

We then apply this to entropy, as a generalisation of \secref{EntanglementOfA2design}.  We go via the 2-norm since the entropy function is not a polynomial.  We find
\begin{theorem}
\label{thm:EntropyBound}
Let $\nu$ be a $4^{-n^2}$-approximate unitary $\frac{n}{10 \log_2 n}$-design on dimension $2^n$ with $n \ge 19$.  Let $d = d_S d_E$ and $2 \le d_S \le 2^{n/10}$ and $\alpha \ge 2$.  Then
\be
\Pr_\nu(S(\psi_S) \le \log_2 d_S - \alpha - \beta) \le 8 \exp_2 \l(-\frac{n}{80\log_2 n} \l( \frac{n}{5} + \alpha \r) \r)
\ee
where $\beta = \frac{1}{\ln 2} \frac{d_S}{d_E}$.
\end{theorem}
We choose a $k$-design for $k=\frac{n}{10 \log_2 n}$ since this is (up to constants) the largest $k$ for which we have an efficient unitary $k$-design construction (see \secref{ApproximateDesigns}).

We then move on to apply our results to ideas in statistical mechanics from \cite{ThermalisationPSW}.  In this paper, the authors show that, for almost all pure states of the universe, any subsystem is very close to the canonical state obtained from the principle of equal a priori probabilities.  This requires the dynamics of the universe to produce a random unitary which would take exponential time in the size of the universe.  We show that the random unitary can be replaced by a $k$-design:
\begin{theorem}
Let $\Omega_S$ be the canonical state of the system and $\rho_S$ be the state after choosing a unitary from an $\eps$-approximate $k$-design.  Let $d_R$ be the dimension of the universe's Hilbert space subject to the arbitrary constraint $R$ (normally this will be a total energy constraint).  Then for $\eps \le \frac{3}{2}\l( \frac{4 d_S^3}{d_R} \r)^{k/8}$, $k \le \frac{4 d_S^2}{9 \pi^3}$
\be
\Pr_\nu( || \rho_S - \Omega_S ||_1 \ge \delta) \le 6 \l(\frac{4 d_S^3}{d_R \delta^2} \r)^{k/8}.
\ee
\end{theorem}

Finally, we use results from \cite{MostStatesUselessMBQCGFE} to show that most states in an $O(1)$-approximate state $n^2$-design on $n$ qubits are useless for measurement based quantum computing, in the sense that any computation using such states could be simulated efficiently on a classical computer.  We do this, following \cite{MostStatesUselessMBQCGFE}, by showing that the states are so entangled that the measurement outcomes are essentially random.

\subsection{Optimality of Results}

An important question is how close our results are to optimal, in terms of their scaling with dimension $d$.  In \thmref{ConcentrationPolynomial}, we will normally have $a=\Theta(d)$ so for $m$ constant, we obtain polynomial bounds, rather than the exponential bounds for full randomness.  This is to be expected:
\begin{theorem}
\label{thm:Optimal}
Let $\nu$ be an $\eps$-approximate unitary $k$-design.  Suppose also that it is discrete i.e.~contains a finite number $S$ of unitaries.  Let $f(U)$ be any function on matrix elements of $U$ and $\mu$ be any constant.  Then either $f(U) = \mu$ for all $U$ in $\nu$ or for some $\delta > 0$
\be
\Pr_\nu(|f-\mu| \ge \delta) \ge p_{min}
\ee
where $p_{min}$ is the probability of choosing the least probable unitary from $\nu$.  If the probability is uniform, $p_{min} = 1/S$.
\end{theorem}
\begin{proof}
There exists at least one $U$ such that $|f(U)-\mu| \ge \delta$ for some $\delta > 0$; the probability of selecting one is at least $p_{min}$.
\end{proof}
\begin{corollary}
Our results are polynomially related to the optimal (i.e.~the optimal bounds can be obtained by raising ours to a constant power).
\end{corollary}
\begin{proof}
Our results apply for any design, so must obey the bound in \thmref{Optimal} for all designs.  The unitary design construction we use (\lemref{kDesignFromExpander}) has $p_{min}=d^{-O(k)}$ hence the bounds cannot scale better than this.
\end{proof}
We can also almost recover the tail bound for full randomness in \thmref{ConcentrationPolynomial}.  Suppose for simplicity that we have an exact design (i.e.~$\eps=0$), so that
\bes
\Pr_\nu(|f - \mu| \ge \delta) \le C \left(\frac{m}{a\delta^2}\right)^m.
\ees
The optimal $m$ is $a \delta^2/e$, which gives
\bes
\Pr_\nu(|f - \mu| \ge \delta) \le C e^{-a \delta^2/e}.
\ees
So our result allows us to interpolate from Markov's inequality, which gives weak bounds, all the way to full Haar randomness and is within a polynomial correction of optimal for the full range.

The remainder of the paper is organised as follows.  In \secref{kDesigns} we formally define $k$-designs and what we mean by approximate designs.  Then in \secref{MainTechnique} we present our main technique for finding large deviation bounds for $k$-designs.  We then apply this to entropy in \secref{Entropy}, to ideas in statistical mechanics in \secref{StatMech} and to using $k$-designs for measurement based quantum computing in \secref{MBQC}.  We then conclude in \secref{Conclusions}.

\section{$k$-designs}
\label{sec:kDesigns}

Here we formally define $k$-designs.
\begin{definition}
\label{def:UnitarykdesignStates}
Let $\nu$ be a distribution on the unitary group.  $\nu$ is a unitary $k$-design if
\begin{equation}
\bbE_\nu \left[ U^{\ot k} \rho \left(U^\dagger\right)^{\ot k} \right] = \bbE_H \l[ U^{\ot k} \rho \left(U^\dagger\right)^{\ot k} \r]
\end{equation}
for all $d^k \times d^k$ complex matrices $\rho$ (not necessarily valid states).
\end{definition}
We can write this as an equivalent, and for our purposes more useful, definition in terms of monomials of the elements of the matrices.  We will first define what we mean by degree of a monomial (or polynomial):
\begin{definition}
A monomial in elements of a matrix $U$ is of degree $(k_1,k_2)$ if it contains $k_1$ conjugated elements and $k_2$ unconjugated elements.  We call it balanced if $k_1=k_2$ and will simply say a balanced monomial has degree $k$ if it is degree $(k,k)$.  A polynomial is of degree $k$ if it is a sum of balanced monomials of degree at most $k$.
\end{definition}
So that, in this definition, $U_{ij} U^*_{pq}$ is a balanced monomial of degree $(1,1)$ and $U_{ij} U_{kl}$ is a monomial of degree $(2,0)$ and is unbalanced.  We now state an equivalent definition of unitary $k$-designs in terms of monomials:
\begin{definition}
\label{def:UnitarykdesignMonomials}
$\nu$ is a unitary $k$-design if, for all balanced monomials $M$ of degree $k$,
\begin{equation}
\bbE_\nu M(U) = \bbE_H M(U).
\end{equation}
\end{definition}
That definitions \ref{def:UnitarykdesignStates} and \ref{def:UnitarykdesignMonomials} are equivalent can be seen by considering matrices $\rho$ of the form $\ket{i_1, i_2, \ldots, i_k} \bra{j_1, j_2, \ldots, j_k}$ in \defref{UnitarykdesignStates}.  Then each element of $U^{\ot k} \rho \left(U^\dagger\right)^{\ot k}$ is a balanced monomial of degree $k$.  Further, each balanced monomial appears for some choice of $\rho$.

We will use state $k$-designs, which are related to unitary $k$-designs although less general:
\begin{definition}
\label{def:StatekdesignStates}
Let $\nu$ be a distribution on states and let $\nu_H$ be the uniform distribution on states, which can be thought of as a random unitary being applied to any fixed state.  Then $\nu$ is a state $k$-design if
\begin{equation}
\bbE_{\ket{\psi} \sim \nu} \l[ \left( \ket{\psi} \bra{\psi} \right)^{\ot k} \r] = \bbE_{\ket{\psi} \sim \nu_H} \l[ \left( \ket{\psi} \bra{\psi} \right)^{\ot k} \r].
\end{equation}
\end{definition}
By considering unitaries acting on a fixed state, it can be seen that a unitary $k$-design can provide a state $k$-design, although the reverse is not necessarily true.

\subsection{Approximate $k$-designs}
\label{sec:ApproximateDesigns}

There are no known efficient constructions of exact unitary $k$-designs.  However, for our purposes, only approximate designs are required.  In \cite{AmbainisEmerson07}, Ambainis and Emerson define an $\eps$-approximate state $k$-design using the $\infty$-norm:
\begin{definition}[\cite{AmbainisEmerson07}]
\label{def:ApproximateStateDesign}
$\nu$ is an $\eps$-approximate state $k$-design if
\begin{equation}
\left|\left|\bbE_{\ket{\psi} \sim \nu} \l[ \left( \ket{\psi} \bra{\psi} \right)^{\ot k} \r] - \bbE_{\ket{\psi} \sim \nu_H} \l[ \left( \ket{\psi} \bra{\psi} \right)^{\ot k} \r]\right|\right|_\infty \le \frac{\eps}{{k+d-1 \choose d-1}}.
\end{equation}
\end{definition}
${k+d-1 \choose d-1}$ appears because it is the dimension of the symmetric subspace.

We will need a definition of an approximate unitary design and will use a slightly different form to the approximate state design definition above that is simpler for our purposes:
\begin{definition}
\label{def:ApproximateUnitarykdesignMonomials}
$\nu$ is an $\eps$-approximate unitary $k$-design if, for all balanced monomials $M$ of degree $\le k$,
\begin{equation}
\label{eq:ApproximateUnitarykdesignMonomials}
\left| \bbE_\nu M(U) - \bbE_H M(U) \right| \le \frac{\eps}{d^k}
\end{equation}
\end{definition}
Finally, we will say a few words about the efficiency of implementing an approximate unitary design.  We would like to be able to have an $\eps$-approximate $k$-design from which we can sample and implement the unitaries using $\poly(\log d, k, \log 1/\eps)$ resources.  Firstly, Ref.~\cite{AmbainisEmerson07} provides an efficient construction of an $\eps$-approximate state $k$-design for all $k \le d/2$.  For unitary designs, we can use the efficient tensor product expander construction from $\cite{EfficientTPE08}$.  A $(d, D, \lambda, k)$ tensor product expander (TPE) is an ensemble of $D$ unitaries $\nu$ in dimension $d$ with, for all $\rho$,
\be
\label{eq:TPEDef}
\l\| \bbE_\nu \l[U^{\ot k} \rho (U^\dagger)^{\ot k}\r] - 
\bbE_H \l[U^{\ot k} \rho (U^\dagger)^{\ot k}\r] 
\right\|_2 \le \lambda \norm{\rho}_2
\ee
where $\lambda < 1$.  In \cite{EfficientTPE08}, an efficient construction is presented with $D$ and $\lambda$ constant for $k = O(\log d/\log \log d)$.  In particular, we can obtain an efficient construction for $k = \frac{\log_2 d}{10\log_2 \log_2 d}$.  To obtain a design according to \defref{ApproximateUnitarykdesignMonomials}, we can iterate the expander:
\begin{lemma}
\label{lem:kDesignFromExpander}
Iterating a $(d, D, \lambda, k)$-TPE $O(k \log d + \log 1/\eps)$ times gives an $\eps$-approximate unitary $k$-design.
\end{lemma}
The slightly technical proof is in the Appendix \ref{sec:kDesignFromExpander}.  Using the efficient TPE construction from \cite{EfficientTPE08}, we have an efficient construction of an $\eps$-approximate $k$-design for $k = O(\log d/\log \log d)$.

\section{Main Technique}
\label{sec:MainTechnique}

The main idea in this paper can be summarised in three steps.  Let $f: \cU(d) \rightarrow \mathbb{C}$ be a polynomial of degree $K$ in the matrix elements of a unitary $U$.  Then to get a concentration bound on $f$ when $U$ is chosen from a $k$-design:
\begin{enumerate}
\item{Find some measure concentration result for $|f(U) - \mu|$ when the unitaries are chosen uniformly at random from the Haar measure.  Normally $\mu$ will be the expectation of $f$.}
\item{Use this to bound the moments $\bbE |f(U) - \mu|^{2m}$ for some integer $m \le \frac{k}{2K}$.}
\item{Then use Markov's inequality and the fact that for a (approximate) $k$-design the moments are (almost) the same as for uniform randomness.  We then optimise the bound for $m$, which will often involve setting $m$ close to the maximum, $\l\lfloor\frac{k}{2K}\r\rfloor$.}
\end{enumerate}
We will now work through each of these steps and finish with a proof of \thmref{ConcentrationPolynomial}.

\subsection{Step 1: Concentration for uniform randomness}
For the first step, we will often start with Levy's Lemma.  This states, roughly speaking, that slowly varying functions in high dimensions are approximately constant.  We quantify `slowly varying' by the Lipschitz constant:
\begin{definition}
The Lipschitz constant $\eta$ (with respect to the Euclidean norm) for a function $f$ is
\begin{equation}
\eta = \min_{U_1, U_2} \frac{| f(U_1) - f(U_2) |}{|| U_1 - U_2 ||_2}.
\end{equation}
\end{definition}
Then we have Levy's lemma:
\begin{lemma}[Levy, see e.g.~\cite{Ledoux}]
\label{lem:Levy}
Let $f$ be an $\eta$-Lipschitz function on $U(d)$ with mean $\bbE f$.  Then
\begin{equation}
\Pr(|f - \bbE f| \ge \delta) \le 4 \exp\left(-\frac{C_1 d \delta^2}{\eta^2}\right)
\end{equation}
where $C_1$ can be taken to be $\frac{2}{9 \pi^3}$.
\end{lemma}

\subsection{Step 2: A bound on the moments}
Levy's Lemma says that $f$ is close to its mean.  This means that $\bbE |f - \bbE f|^m$ should be small.  We will bound the moments for slightly more general concentration results:
\begin{lemma}
\label{lem:MomentBoundFromLargeDeviationBound}
Let $X$ be any random variable with probability concentration
\begin{equation}
\label{eq:LargeDeviationAssumption}
\Pr(|X - \mu| \ge \delta) \le C e^{-a \delta^2}.
\end{equation}
(Normally $\mu$ will be the expectation of $X$, although the bound does not assume this.)  Then
\begin{equation}
\bbE |X - \mu|^m \le C \Gamma(m/2+1) a^{-m/2} \le C \left( \frac{m}{2a} \right)^{m/2}
\end{equation}
for any $m > 0$.
\end{lemma}
\begin{proof}
This proof is based on the proof of an analogous result by Bellare and Rompel \cite{BellareRompel}, Lemma A.1.

Note that, for any random variable $Y \ge 0$, 
\begin{equation}
\bbE Y = \int_0^\infty \Pr(Y \ge y) dy.
\end{equation}
Therefore
\bas
\bbE |X-\mu|^m &= \int_0^\infty \Pr(|X-\mu|^m \ge x) dx \\
&= \int_0^\infty \Pr(|X-\mu| \ge x^{1/m}) dx \\
&\le C \int_0^{\infty} \exp(-a x^{2/m}) dx
\eas
where in the last line we used the assumed large deviation bound \eq{LargeDeviationAssumption}.  To evaluate this integral, use the change of variables $y = a x^{2/m}$ to get
\bas
\bbE |X-\mu|^m &\le \frac{Cm}{2} a^{-m/2} \int_0^\infty e^{-y} y^{m/2-1} dy\\
&= C a^{-m/2} \Gamma(m/2+1) \\
&\le C \left( \frac{m}{2a} \right)^{m/2}.\qedhere
\eas
\end{proof}

\subsection{Step 3: A concentration bound for a $k$-design}

Now we show how to obtain a measure concentration result for polynomials when the unitaries are selected from an approximate $k$-design.  We first show that the moments of $|f-\mu|$ for $f$ a polynomial are close to the Haar measure moments:
\begin{lemma}
\label{lem:ApproxkdesignMoments}
Let $f$ be a polynomial of degree $K$ and $\mu$ be any constant.  Let $f = \sum_{i=1}^t \alpha_i M_i$ where each $M_i$ is a monomial.  Let $\alpha(f) = \sum_i |\alpha_i|$.  Then for $m$ an integer with $2mK \le k$ and $\nu$ an $\eps$-approximate $k$-design,
\be
\bbE_\nu |f - \mu|^{2m} \le \bbE_H | f-\mu |^{2m} + \frac{\eps}{d^k} \left(\alpha + | \mu | \right)^{2m}.
\ee
\end{lemma}
\begin{proof}
For simplicity, we assume that $f$ and $\mu$ are real.  Our proof easily generalises to the complex case.

Firstly we calculate $|\bbE_\nu f^i - \bbE_H f^i|$ using the multinomial theorem:
\bas
|\bbE_\nu f^i - \bbE_H f^i| &=
\l| \sum_{k_1 + \ldots + k_t = i} {i \choose k_1, \ldots, k_t} \alpha_1^{k_1} \ldots \alpha_t^{k_t} \l( \bbE_\nu M_1^{k_1} \ldots M_t^{k_t} - \bbE_H M_1^{k_1} \ldots M_t^{k_t}\r) \r| \\
&\le \sum_{k_1 + \ldots + k_t = i} {i \choose k_1, \ldots, k_t} |\alpha_1|^{k_1} \ldots |\alpha_t|^{k_t} \l| \bbE_\nu M_1^{k_1} \ldots M_t^{k_t} - \bbE_H M_1^{k_1} \ldots M_t^{k_t} \r| \\
&\le \frac{\eps}{d^k} \sum_{k_1 + \ldots + k_t = i} {i \choose k_1, \ldots, k_t} |\alpha_1|^{k_1} \ldots |\alpha_t|^{k_t} \\
&= \frac{\eps}{d^k} \alpha^i.
\eas
We now calculate $\bbE_\nu |f - \mu|^{2m}$:
\bas
\l| \bbE_\nu | f - \mu |^{2m} - \bbE_H | f - \mu |^{2m} \r|
&= \l| \bbE_\nu (f - \mu)^{2m} - \bbE_H (f - \mu)^{2m} \r| \\
&= \l|\sum_{i = 0}^{2m} {2m \choose i} (\bbE_\nu f^i - \bbE_H f^i) (-\mu)^{2m-i} \r|\\
&\le \sum_{i = 0}^{2m} {2m \choose i} |\bbE_\nu f^i - \bbE_H f^i| |\mu|^{2m-i}\\
&\le \frac{\eps}{d^k} \sum_{i = 0}^{2m} {2m \choose i} \alpha^i |\mu|^{2m-i}\\
&= \frac{\eps}{d^k} \left(\alpha + | \mu | \right)^{2m}.\qedhere
\eas
\end{proof}

Now we can simply apply Markov's inequality to prove \thmref{ConcentrationPolynomial}.
\begin{proof}[Proof of \thmref{ConcentrationPolynomial}]
Apply Markov's inequality and Lemmas \ref{lem:MomentBoundFromLargeDeviationBound} and \ref{lem:ApproxkdesignMoments}:
\bas
\Pr_\nu(|f - \mu| \ge \delta) &= \Pr_\nu(|f - \mu|^{2m} \ge \delta^{2m}) \\
&\le \frac{\bbE_\nu | f - \mu |^{2m}}{\delta^{2m}} \\
&\le \frac{1}{\delta^{2m}} \left( C \left(\frac{m}{a}\right)^m + \frac{\eps}{d^k} \left(\alpha + | \mu | \right)^{2m}\right).\qedhere
\eas 
\end{proof}

We finish this section with two remarks.  Firstly, provided $\alpha(f)$ (the sum of the absolute value of all the coefficients) is at most polynomially large in $d$, we can choose $\eps$ to be polynomially small to cancel this at no change to the asymptotic efficiency.  Secondly, when applying the theorem we will optimise the choice of $m$ (and normally choose $k = 2mK$).  Often $a=\Theta(d)$ and the optimal choice of $m$ is often $\Theta(d)$ as well.  However, we will not take $m$ so large because we can only implement an efficient $k$-design for $k = O(\log d/\log \log d)$.

\section{Application 1: Entropy of a $k$-design}
\label{sec:Entropy}

We now apply the above to show that most unitaries in a $k$-design generate large amounts of entropy across any bipartition, provided the dimensions are sufficiently far apart.  This means that, for any initial state, for most choices of a unitary from a $k$-design applied to the state, the resulting state will be highly entangled.  We go via the purity of the reduced density matrix, since the entropy function is not a polynomial.

We will call the two systems $S$ (the `system') and $E$ (the `environment') and calculate the purity of the reduced state.  That the purity, $\tr\left[\left( \tr_E U \rho U^\dagger \right)^2\right]$, is a polynomial of degree 2 is easily seen by noting that the trace is linear and the reduced state is squared.  However, we should check that there are not too many terms or terms with large coefficients.  To do this, we should calculate $\alpha$ to apply \thmref{ConcentrationPolynomial}.

There is a general method for calculating $\alpha(f)$ which we will use.  Write $f(U) = \sum_i \alpha_i M_i(U)$ for monomials $M_i$.  To evaluate $\alpha(f) = \sum_i | \alpha_i |$, calculate $f(A)$ where $A$ is the matrix with all entries equal to $1$ (so that $M_i(A) = 1$) and replace $\alpha_i$ with $| \alpha_i|$. Using this here we find
\bas
\alpha &= d^2\l(\sum_{i j} | \rho_{ij} | \r)^2 \\
&\le d^4 \sum_{ij} | \rho_{ij} |^2 \\
&= d^4 || \rho ||_2^2 \\
&\le d^4.
\eas


We now calculate the expected purity:
\begin{lemma}
\label{lem:ExpectedPurity}
The expected purity of the reduced state is $\frac{d_S + d_E}{d+1}$, where $d_S$ is the dimension of subsystem $S$ and $d_E = d/d_S$ is the dimension of subsystem $E$.
\end{lemma}
\begin{proof}
We have
\be
\bbE_H || \psi_S ||_2^2 = \bbE_H \l[ \tr \cF_{S_1 S_2} (\tr_E U \rho U^\dagger \ot \tr_E U \rho U^\dagger) \r]
\ee
where $\cF_{S_1 S_2}$ is swap acting between systems $S_1$ and $S_2$.  By linearity of the trace, we can commute the $\bbE_H$ through and use $\bbE_H \l[ U \rho U^\dagger \ot U \rho U^\dagger \r] = \frac{I_{12} + \cF_{12}}{d(d+1)}$ to find
\bas
\bbE_H || \psi_S ||_2^2 &= \tr \l[\frac{\cF_{S_1 S_2}}{d(d+1)} (d_E^2 I_{S_1 S_2} + d_E \cF_{S_1 S_2}) \r] \\
&= \frac{1}{d(d+1)} (d_E^2 d_S + d_E d_S^2) \\
&= \frac{d_S + d_E}{d+1}\qedhere
\eas
\end{proof}

Working out the higher moments in this way is difficult so we use Levy's Lemma and \lemref{MomentBoundFromLargeDeviationBound}.  To use Levy's Lemma, all we have to do is find the Lipschitz constant for the purity:
\begin{lemma}
The Lipschitz constant for purity is $\le 2$.
\end{lemma}
\begin{proof}
\bas
\eta &= \sup_{\psi, \phi} \frac{\left| ||\psi_S||_2^2 - ||\phi_S||_2^2 \right|}{||\psi - \phi||_2} \\
&= \sup_{\psi, \phi} \frac{\left| ||\psi_S||_2 - ||\phi_S||_2 \right| (||\psi_S||_2 + ||\phi_S||_2)}{||\psi - \phi||_2}
\eas
Now we use $\left| ||S||_2 - ||T||_2 \right| \le ||S-T||_2$ to find
\bes
\eta \le \sup_{\psi, \phi} (||\psi_S||_2 + ||\phi_S||_2) \le 2
\ees
using the fact that the purity is upper bounded by 1.
\end{proof}

\begin{lemma}
\label{lem:MessyEntropyBound}
For $\mu = \frac{d_S + d_E}{d+1}$ and $m$ an integer with $m \le k/4$ and $\nu$ an $\eps$-approximate $k$-design,
\be
\label{eq:MessyEntropyBound}
\Pr_\nu(S(\psi_S) \le -\log_2 \mu - \alpha) \le \frac{1}{(\mu (2^\alpha-1))^{2m}} \left(4 \left(\frac{4 m}{C_1 d}\right)^m + \frac{\eps}{d^k}(d^4 + \mu)^{2m} \right).
\ee
\end{lemma}
\begin{proof}
We use the fact that von Neumann entropy is lower bounded by the Renyi 2-entropy i.e.~$-\log_2 || \psi_S ||_2^2$:
\begin{equation}
S(\psi_S) \ge S_2(\psi_S) = - \log_2 || \psi_S ||_2^2.
\end{equation}
Then
\bas
\Pr_\nu(S(\psi_S) \le - \log_2 (1+\delta) \mu) &\le \Pr_\nu(S_2(\psi_S) \le - \log_2 (1+\delta) \mu) \\
&= \Pr_\nu(|| \psi_S ||_2 \ge (1+\delta)\mu) \\
&\le \Pr_\nu(\left| || \psi_S ||_2 - \mu \right|\ge \delta\mu) \\
&\le \frac{1}{(\mu \delta)^{2m}} \left(4 \left(\frac{4 m}{C_1 d}\right)^m + \frac{\eps}{d^k}(d^4 + \mu)^{2m} \right)
\eas
using \thmref{ConcentrationPolynomial} in the last line.
\end{proof}
We have written this in a more convenient form in \thmref{EntropyBound} which is proved in the Appendix \ref{sec:EntropyBoundProof}.  This is to be compared with the fully random version \thmref{EntropyTailBoundFullRandomness}.  As expected, we have $n = \log_2 d$ appearing in the exponent rather than $d$.  Note also that our bound does not work well for $d_S \approx d_E$.  In fact, in this case, we do not get a bound that improves with dimension.  To get this, $d_S$ must be polynomially smaller than $d_E$.

\section{Application 2: $k$-designs and Statistical Mechanics}
\label{sec:StatMech}

We can also apply these ideas to derandomise some of the arguments on the foundations of statistical mechanics in \cite{ThermalisationPSW}.  In this paper, the authors develop the idea that the uncertainty in statistical mechanics comes from entanglement rather than the traditional assumption of the principle of equal a priori probabilities.  They consider the universe being in a pure quantum state and that the uncertainty in the state of a subsystem comes from the entanglement between this system and the rest of the universe.

The setting is that there is an arbitrary global constraint $R$.  Often this will be a total energy constraint although this is not assumed.  Let the Hilbert space of states satisfying $R$ be $\cH_R$.  Then let the system and environment Hilbert spaces be $\cH_S$ and $\cH_E$ respectively.  Then
\be
\cH_R \subseteq \cH_S \ot \cH_E.
\ee
Let the dimensions be $d_R$, $d_S$ and $d_E$ and let $\cE_R = \frac{I_R}{d_R}$.  Note that $d_R \le d_S d_E$, unlike in the above where we took $d = d_S d_E$.  Normally we will have $d_S \ll d_R$.  The principle of equal a priori probabilities says that the state of the universe is $\cE_R$ which implies the subsystem state is the canonical state, given by
\be
\label{eq:CanonicalState}
\Omega_S = \tr_E(\cE_R).
\ee
The main result of \cite{ThermalisationPSW} (the `principle of \emph{apparently} equal a priori probabilities') is that, for almost all pure states of the universe, the subsystem state is almost exactly the canonical state.  
\begin{theorem}[Theorem 1 of \cite{ThermalisationPSW}]
\label{thm:ThermalisationPSW}
For a randomly chosen state $\ket{\phi} \in \mathcal{H}_R \subseteq \mathcal{H}_S \ot \mathcal{H}_E$ and arbitrary $\eps > 0$, the distance between the reduced density matrix of the system $\rho_S = \tr_E(\ket{\phi} \bra{\phi})$ and the canonical state $\Omega_S$ (\eq{CanonicalState}) is given probabilistically by
\begin{equation}
\Pr_H\left( || \rho_S - \Omega_S ||_1 \ge \eps + \sqrt{\frac{d_S}{d_E^{\text{\rm eff}}}}\right) \le 2 \exp\left(-C_2 d_R \eps^2\right)
\end{equation}
where $C_2 = 1/(18 \pi^3)$ and $d_E^{\text{\rm eff}} = \frac{1}{\tr \Omega_E^2} \ge \frac{d_R}{d_S}$.
\end{theorem}

This result gives compelling evidence to replace the principle of equal a priori probabilities with the principle of apparently equal a priori probabilities, but it does not address the problem of how the system reaches this state.  It will take an extremely (exponentially) long time for the universe to reach a random pure state, in contrast to the observed fact that thermalisation occurs quickly.  Here, we show that for almost all unitaries in a $k$-design applied to the universe, the subsystem state is close to the canonical state.  Since these unitaries can be implemented and sampled from efficiently, this means that equilibrium could be reached quickly to match observations.

We are now ready to show that a $k$-design gives a small $|| \rho_S - \Omega_S ||_1$.  First, we have to modify \lemref{MomentBoundFromLargeDeviationBound} slightly:
\begin{lemma}
\label{lem:MomentBoundFromLargeDeviationBound2}
Let $X$ be any non-negative random variable with probability concentration
\begin{equation}
\label{eq:LargeDeviationAssumption2}
\Pr(X \ge \delta + \eta) \le C e^{-a \delta^2}
\end{equation}
where $\eta \ge 0$.  Then
\begin{equation}
\bbE X^m \le C \left( \frac{2m}{a} \right)^{m/2} + (2\eta)^m
\end{equation}
for any $m > 0$.
\end{lemma}
\comment{\begin{proof}
For $\delta \ge \eta$, $\Pr(X \ge 2\delta) \le C e^{-a \delta^2}$.  Alternatively, if $\delta \ge 2\eta$, $\Pr(X \ge \delta) \le C e^{-a \delta^2/4}$.  Now
\begin{align}
\Expect X^m &= \int_0^\infty \Pr(X^m \ge x) dx \\
&= \int_0^\infty \Pr(X \ge x^{1/m}) dx \\
&= \int_{(2\eta)^m}^\infty \Pr(X \ge x^{1/m}) dx + \int_0^{(2\eta)^m} \Pr(X \ge x^{1/m}) dx \\
&\le \int_{(2\eta)^m}^\infty C \exp(-a x^{2/m}/4) dx + (2\eta)^m \\
&\le \int_0^\infty C \exp(-a x^{2/m}/4) dx + (2\eta)^m \\
&\le C \left( \frac{2m}{a} \right)^{m/2} + (2\eta)^m
\end{align}
where the last line follows from evaluating the integral, as in the proof of \lemref{MomentBoundFromLargeDeviationBound}.
\end{proof}}
The proof is very similar to the proof of \lemref{MomentBoundFromLargeDeviationBound}.

Now we state and prove the main result in this section:
\begin{theorem}
Let $\nu$ be an $\eps$-approximate unitary $k$-design.  Then
\be
\label{eq:ThermalisationMessy}
\Pr_\nu( || \rho_S - \Omega_S ||_1 \ge \delta) \le \l(\frac{d_S}{\delta^2}\r)^{k/8} \l( 2 \l(\frac{k}{2 C_2 d_R}\r)^{k/8} + \l( \frac{4 d_S^2}{d_R} \r)^{k/8} + \frac{\eps}{d_R^k}(d_R^2+1)^{4m}\r).
\ee
In particular, with $\eps = \frac{3}{2}\l( \frac{4 d_S^3}{d_R} \r)^{k/8}$, $k \le 8C_2 d_S^2$,
\be
\label{eq:ThermalisationSimplified}
\Pr_\nu( || \rho_S - \Omega_S ||_1 \ge \delta) \le 6 \l(\frac{4 d_S^3}{d_R \delta^2} \r)^{k/8}.
\ee
\end{theorem}
Again, we need $d_S$ to be polynomially smaller than $d_R$ to obtain non-trivial bounds.

\begin{proof}
We go via the 2-norm and use Lemmas \ref{lem:MomentBoundFromLargeDeviationBound2} and \ref{lem:ApproxkdesignMoments}.

We have from \thmref{ThermalisationPSW} that
\begin{equation}
\Pr_H(||\rho_S - \Omega_S||_1 \ge \delta + \eta) \le 2 e^{-C_2 d_R \delta^2}
\end{equation}
where $\eta = \sqrt{\frac{d_S}{d_E^\text{eff}}} \le \frac{d_S}{\sqrt{d_R}}$.  Since $||\rho_S - \Omega_S||_2 \le ||\rho_S - \Omega_S||_1$,
\begin{equation}
\Pr_H(||\rho_S - \Omega_S||_2 \ge \delta + \eta) \le 2 e^{-C_2 d_R \delta^2}.
\end{equation}
We now apply \lemref{MomentBoundFromLargeDeviationBound2} to get
\begin{equation}
\Expect_H ||\rho_S - \Omega_S||_2^{2m} \le 2\left(\frac{4m}{C_2 d_R}\right)^{m} + (2 \eta)^{2m}.
\end{equation}
So for $m \le k/4$, using Markov's inequality and \lemref{ApproxkdesignMoments} (with $\mu = 0$) on the polynomial $||\rho_S - \Omega_S||_2^{2}$ :
\be
\Pr_\nu (||\rho_S - \Omega_S||_2 \ge \delta) \le \frac{1}{\delta^{2m}} \left( 2\left(\frac{4m}{C_2 d_R}\right)^{m} + (2 \eta)^{2m} + \frac{\eps}{d_R^k}(d_R^2+1)^{4m} \r).
\ee
Here, we used an estimate of $\alpha$, the sum of the moduli of the coefficients:
\be
\alpha \le (d_R^2+1)^2
\ee
which we obtain via a similar calculation to that in \secref{Entropy}.

Now we go back to the 1-norm, using $||\rho_S - \Omega_S||_1 \le \sqrt{d_S} ||\rho_S - \Omega_S||_2$ to get
\begin{align}
\Pr_\nu (||\rho_S - \Omega_S||_1 \ge \delta) &\le \Pr_\nu (||\rho_S - \Omega_S||_2 \ge \delta/\sqrt{d_S}) \\
&\le \l(\frac{d_S}{\delta^2}\r)^{m} \l( 2 \l(\frac{4m}{C_2 d_R}\r)^{m} + \l( 2 \eta \r)^{2m} + \frac{\eps}{d_R^k}(d_R^2+1)^{4m} \r).
\end{align}
To obtain the result in \eq{ThermalisationMessy}, we just use $\eta \le \frac{d_S}{\sqrt{d_R}}$.

To prove the simplified version, first use, as in \secref{Entropy}, that $(d_R^2+1)^{4m} \le 2 d_R^{8m}$ for $m \le d_R^2/8$.  This is implied by $k \le 8C_2 d_S^2$.  We then set $m = k/8$ to find
\be
\Pr_\nu( || \rho_S - \Omega_S ||_1 \ge \delta) \le 2 \l(\frac{k d_S}{2 C_2 d_R \delta^2}\r)^{k/8} + \l( \frac{4 d_S^3}{d_R \delta^2} \r)^{k/8} + 2\frac{\eps}{\delta^{k/4}}.
\ee
Then, using $k \le 8 C_2 d_S^2$, with $\eps \le \frac{3}{2}\l( \frac{4 d_S^3}{d_R} \r)^{k/8}$, we obtain the simplified result \eq{ThermalisationSimplified}.
\end{proof}

\section{Application 3: Using $k$-designs for Measurement Based Quantum Computing}
\label{sec:MBQC}

Here we apply our ideas to derandomise some results of Gross, Flammia and Eisert in \cite{MostStatesUselessMBQCGFE} and Bremner, Mora and Winter in \cite{MostStatesUselessMBQCBMW}.  The main result in these two papers is that most states do not offer any advantage over classical computation when used in the measurement based quantum computing (MBQC) model.  In MBQC, a classical computer is given access to a large quantum state on which it can do single qubit measurements.  Some states allow for universal quantum computation whereas others do not add any extra power to the classical computer.  These results are concerned with the question of characterising which states do and do not work.  Showing that random states do not give any speed up shows that useful states for MBQC are not generic and so must be carefully constructed.

While the results in these two papers are similar, we will concentrate on the methods from \cite{MostStatesUselessMBQCGFE} since their methods are simpler to apply here.  They prove their result by showing that most states are very entangled in the geometric measure (see \defref{GeometricMeasure}).  They then use this to show that the measurement outcomes of even the best possible measurement scheme are almost completely random.  In fact, the state could be thrown away and the measurement outcomes replaced with random numbers to solve the computational problem just as efficiently.  This shows that you can classically simulate any quantum computation that uses these highly entangled states.  The measure of entanglement they use is the geometric measure:
\begin{definition}
\label{def:GeometricMeasure}
The geometric measure of entanglement of a state $\ket{\Psi}$ is \cite{GeometricEntanglementShimony, BarnumLinden01}
\be
E_g(\ket{\Psi}) = -\log_2 \sup_{\alpha \in \cP} | \braket{\alpha}{\Psi} |^2.
\ee
where $\cP$ is the set of all product states.
\end{definition}
They show that any MBQC using a state $\ket{\Psi}$ with $E_g(\ket{\Psi}) = n - O(\log_2 n)$ can be efficiently simulated classically.  They then show that (we abuse notation slightly by writing $\Pr_H$ for $\Pr_{\ket{\Psi} \sim \nu_H}$, etc.)
\begin{theorem}[\cite{MostStatesUselessMBQCGFE}, Theorem 2]
\label{thm:HaarRandomLargeGeometricEntropy}
For $n \ge 11$,
\be
\Pr_H (E_g(\ket{\Psi}) \le n - 2 \log_2 n - 3) \le e^{-n^2}.
\ee
\end{theorem}
This shows that most states are useless.  We derandomise this result to show that most states in an $\eps$-approximate ($\eps$ can be taken as a constant) state $n^2$-design have high geometric measure of entanglement and thus are useless in the same way.

We could apply our technique and use \thmref{ConcentrationPolynomial} but in this case, it is simpler to directly bound the probability using Markov's inequality.
\begin{lemma}
\label{lem:RandomStateOverlap}
\begin{equation}
\Pr_\nu(| \braket{\Phi}{\Psi} |^2 \ge \delta) \le (1+\eps)\frac{m!}{(d\delta)^m} \le (1+\eps)\left(\frac{m}{d\delta}\right)^m
\end{equation}
where $\ket{\Psi}$ is chosen from an $\eps$-approximate state $k$-design $\nu$, $m \le k$ and a positive integer and $\ket{\Phi}$ is any fixed state.
\end{lemma}
\begin{proof}
We prove this bound directly using Markov's inequality:
\bas
\Pr_\nu(| \braket{\Phi}{\Psi} |^2 \ge \delta) &= \Pr_\nu(| \braket{\Phi}{\Psi} |^{2m} \ge \delta^m) \\
&\le \frac{\bbE_\nu  | \braket{\Phi}{\Psi} |^{2m}}{\delta^m} \\
&= \frac{\bbE_\nu  \bra{\Phi}^{\ot m} \ket{\Psi}^{\ot m} \bra{\Psi}^{\ot m} \ket{\Phi}^{\ot m}}{\delta^m} \\
&= \frac{\bra{\Phi}^{\ot m} \bbE_\nu \l[ \ket{\Psi}^{\ot m} \bra{\Psi}^{\ot m} \r] \ket{\Phi}^{\ot m}}{\delta^m} \\
&\le \frac{\bra{\Phi}^{\ot m} (1+\eps)\frac{\Pi^\text{sym}_m}{{m+d-1 \choose d-1}} \ket{\Phi}^{\ot m}}{\delta^m} \\
&= \frac{1+\eps}{{m+d-1 \choose d-1} \delta^m} \\
&\le \frac{(1+\eps)m!}{(d \delta)^m} \le (1+\eps)\left(\frac{m}{d \delta} \right)^m.\qedhere
\eas
\end{proof}
We now prove the main result in this section:
\begin{theorem}
\label{thm:LargeGeometricEntanglement}
For $\ket{\Psi}$ randomly drawn from an $\eps$-approximate state $k$-design with $d = 2^n$
\begin{equation}
\Pr_\nu(E_g(\ket{\Psi}) \le n - \delta) \le (1+\eps) \exp_2 (k \log_2 2k + 4n \log_2 10n - k\delta + 4n(n-\delta)).
\end{equation}
In particular, for $k = n^2$, $\delta = 3 \log_2 n + 5$ and $\eps = 1$,
\begin{equation}
\Pr_\nu(E_g(\ket{\Psi}) \le n - 3 \log_2 n - 5) \le 2 \cdot n^{-n^2}.
\end{equation}
\end{theorem}
We note that this bound is almost the same as in \thmref{HaarRandomLargeGeometricEntropy}.  It only works for slightly larger deviations from $n$, which is why we obtain a slightly better probability bound.  Note also that we can obtain an exponential bound in $n$ (not $d=2^n$) because the design is exponentially large in $n$.
\begin{proof}
This proof closely mirrors the proof of Theorem 2 in \cite{MostStatesUselessMBQCGFE}.  We use the idea of a $\gamma$-net.  $\cN_{\gamma, n}$ is a $\gamma$-net on product states if
\be
\sup_{\ket{\alpha} \in \cP} \inf_{\ket{\tilde{\alpha}} \in \cN_{\delta, n}} \big|\big| \ket{\alpha} - \ket{\tilde{\alpha}} \big|\big|_2 \le \gamma/2.
\ee
In \cite{MostStatesUselessMBQCGFE}, it is shown that such a net exists with $| \mathcal{N}_{\gamma, n}| \le (5n/\gamma)^{4n}$.  We then proceed by showing that most states in the state design have small overlap with every state in the net using the union bound and \lemref{RandomStateOverlap}.  Finally, since every state is close to one in the net, we can show that most states in the design have small overlap with every product state.

We now formalise the above.  Using \lemref{RandomStateOverlap} and the union bound,
\begin{equation}
\label{eq:NetStateOverlap}
\Pr_\nu \left( \sup_{\ket{\tilde{\alpha}} \in \mathcal{N}_{\gamma, n}} | \braket{\tilde{\alpha}}{\Psi} |^2 \ge \delta'/2 \right) \le | \mathcal{N}_{\gamma,n} | (1+\eps) \left(\frac{2k}{d\delta'}\right)^k \le \left(\frac{5n}{\gamma}\right)^{4n} (1+\eps) \left(\frac{2k}{2^n \delta'}\right)^k.
\end{equation}
Now, we need to bound
\bas
\Pr_\nu(E_g(\ket{\Psi}) \le n - \delta) &= \Pr_\nu\left(-\log_2 \sup_{\ket{\alpha} \in \mathcal{P}} | \braket{\alpha}{\Psi} |^2 \le n - \delta\right) \\
&= \Pr_\nu\left(\sup_{\ket{\alpha} \in \mathcal{P}} | \braket{\alpha}{\Psi} |^2 \ge 2^{-(n - \delta)}\right).
\eas
We now claim that
\be
\label{eq:OverlapInNet}
\sup_{\ket{\alpha} \in \cP} | \braket{\alpha}{\Psi} |^2 \ge \delta' \Rightarrow \sup_{\ket{\tilde{\alpha}} \in \cN_{\delta'/2, n}} | \braket{\tilde{\alpha}}{\Psi} |^2 \ge \delta'/2.
\ee
To prove this claim, let $\ket{\alpha}$ be the state that achieves the supremum on the left hand side, and let $\ket{\tilde{\alpha}}$ be the state closest to it in the $\delta'/2$-net.  It is shown in \cite{MostStatesUselessMBQCGFE} that this implies for any $\ket{\Psi}$
\be
\left| | \braket{\alpha}{\Psi} | ^2 - | \braket{\tilde{\alpha}}{\Psi} | ^2 \right| \le \delta'/2.
\ee
Therefore
\bas
| \braket{\tilde{\alpha}}{\Psi} |^2 &\ge  | \braket{\alpha}{\Psi} |^2 - \delta'/2 \\
&\ge \delta'/2.
\eas
This implies that the supremum over all states in the net must be at least $\delta'/2$ to prove the claim.

We can now finish the proof.  Set $\delta' = 2^{-(n - \delta)}$ in \eq{OverlapInNet} and use \eq{NetStateOverlap} with $\gamma = \delta'/2$ to find
\bas
\Pr_\nu\left(\sup_{\ket{\alpha} \in \mathcal{P}} | \braket{\alpha}{\Psi} |^2 \ge 2^{-(n - \delta)}\right)
&\le \Pr_\nu\left(\sup_{\ket{\tilde{\alpha}} \in \mathcal{N}_{2^{-(n-\delta)-1}, n}} | \braket{\tilde{\alpha}}{\Psi} |^2 \ge 2^{-(n - \delta)-1}\right) \\
&\le (1+\eps) \exp_2 (k \log_2 2k + 4n \log_2 10n - k\delta + 4n(n-\delta)).\qedhere
\eas
\end{proof}
Combining this with the arguments of \cite{MostStatesUselessMBQCGFE} shows that most states in a state $n^2$-design on $n$ qubits are useless for MBQC.  This shows that even many efficiently preparable states are useless.

\section{Conclusions}
\label{sec:Conclusions}

We have seen how to turn large deviation bounds for Haar-random unitaries into bounds for $k$-designs.  The main technique was applied to show that unitaries from $k$-designs generate large amounts of entanglement.  Then we showed that, if the dynamics of the universe produced a $k$-design, the entanglement generated would be sufficient to reproduce the principle of equal a priori probabilities.  Finally we showed that most states in sufficiently large state designs are useless for measurement based quantum computing, in the sense that computation using them can be efficiently simulated classically.

However, there are other bounds for which our technique does not work.  Since we cannot obtain exponential bounds for polynomially sized designs, our technique cannot directly derandomise some bounds.  Some results, for example showing that the $\infty$-norm of the reduced state of a random pure state is close to $1/d_S$ \cite{SuperdenseCodingHHL}, are proven by using an $\eps$-net of states and the union bound.  Since the $\eps$-net is exponentially large, exponentially small bounds are required.  We do not know how to apply our idea to results of this kind and still have $k=\poly(\log d)$.  (Note that we could cope with the $\eps$-net in \secref{MBQC} since it was just a net on product states which is considerably smaller.)

It is also possible that our ideas could be used to completely derandomise some constructions (e.g.~locking \cite{RandomizingQuantumStates04, Locking04}).  If we could show that unitaries drawn from a $k$-design work with non-zero probability, and come up with an efficient sampling method, then we could obtain efficient randomised constructions.

{\bf Acknowledgements.} I am grateful for funding from the U.K. Engineering and Physical Science Research Council through ``QIP IRC.''   I thank Aram Harrow for many useful discussions on this topic, comments on earlier drafts of this manuscript and for suggesting the use of existing large deviation bounds to bound the high moments.  I also thank Toby Cubitt for suggesting applying this method to the results of \cite{ThermalisationPSW}, Ashley Montanaro for useful discussions and comments on drafts of this manuscript as well as Andreas Winter and Michael Bremner for useful discussions and comments.

\appendix

\section{Appendix}

Here we present some miscellaneous proofs.

\subsection{Proof of Lemma 2.7}
\label{sec:kDesignFromExpander}

\begin{proof}[Proof of \lemref{kDesignFromExpander}]
We claim that, if for all $d^k \times d^k$ matrices $\rho$,
\be
\label{eq:2NormError}
\l\| \bbE_{\sigma} \l[U^{\ot k} \rho (U^\dagger)^{\ot k}\r] - 
\bbE_H \l[U^{\ot k} \rho (U^\dagger)^{\ot k}\r] 
\right\|_2 \le \frac{\eps}{d^{3k/2}} \norm{\rho}_2
\ee
then $\sigma$ is an $\eps$-approximate $k$-design.  To prove this claim, let $m \le k$ and take $M$ to be any balanced monomial of degree $m$.  Write $M = U_{p_1 q_1} \ldots U_{p_m q_m} U_{r_1 s_1}^* \ldots U_{r_m s_m}^*$.  Then let $\rho_m = \ket{q_1, \ldots, q_m}\bra{s_1, \ldots, s_m}$.  Let $\cE_{\sigma,k}(\rho) = \bbE_{\sigma} \l[ U^{\ot k} \rho \left(U^\dagger\right)^{\ot k} \r]$, $\cE_{H,k}(\rho) = \bbE_H \l[ U^{\ot k} \rho \left(U^\dagger\right)^{\ot k} \r]$ and $\rho_k = \rho_m \ot \frac{I^{\ot k-m}}{d^{k-m}}$.  Then
\bas
\frac{\eps}{d^{3k/2}} &\ge \l|\l| \cE_{\sigma,k}(\rho_k) - \cE_{H,k}(\rho_k) \r|\r|_2 \\
&= \l|\l| \l(\cE_{\sigma,m}(\rho_m) - \cE_{H,m}(\rho_m)\r) \ot \frac{I^{\ot k-m}}{d^{k-m}} \r|\r|_2 \\
&= \frac{1}{\sqrt{d^{k-m}}} || \cE_{\sigma,m}(\rho_m) - \cE_{H,m}(\rho_m) ||_2 \\
\eas
We then use the fact that the largest matrix element is upper bounded by the 2-norm.  For any matrix $A$,
\bas
|A_{ij}| \le \sqrt{\sum_{i'j'} |A_{i'j'}|^2} = \sqrt{\tr A^\dagger A} = ||A||_2.
\eas
For us, this implies
\be
| (\cE_{\sigma,m}(\rho_m) - \cE_{H,m}(\rho_m))_{p_1 \ldots p_m, r_1 \ldots r_m} | \le || \cE_{\sigma,m}(\rho_m) - \cE_{H,m}(\rho_m) ||_2
\ee
which gives
\be
|\bbE_\nu M - \bbE_H M | \le \frac{\eps}{d^k}
\ee
to prove the claim.

Then we just have to show how to obtain \eq{2NormError} from \eq{TPEDef}.  Iterating the TPE $t$ times gives
\be
\l\| \bbE_{\nu^t} \l[U^{\ot k} \rho (U^\dagger)^{\ot k}\r] - 
\bbE_H \l[U^{\ot k} \rho (U^\dagger)^{\ot k}\r] 
\right\|_2 \le \lambda^t
\ee
where $\nu^t$ is the ensemble obtained by applying $t$ unitaries from $\nu$.  Now choose $t$ such that $\lambda^t \le \frac{\eps}{d^{3k/2}}.$
\end{proof}

\subsection{Proof of Theorem 1.3}
\label{sec:EntropyBoundProof}

Here we prove the more convenient form of \lemref{MessyEntropyBound} stated as \thmref{EntropyBound}.

\begin{proof}[Proof of \thmref{EntropyBound}]

Firstly, we will write the left hand side of \eq{MessyEntropyBound} in a more useful way.  Using $\ln(1+x) \le x$, we find
\bes
-\log_2 \mu \ge \log_2 d_S - \beta
\ees
where $\beta = \frac{1}{\ln 2} \frac{d_S}{d_E}$, following the notation in \cite{AspectsOfGenericEntanglement}.  This means
\bas
\Pr_\nu(S(\psi_S) \le \log_2 d_S - \alpha - \beta) &\le \Pr_\nu(S(\psi_S) \le -\log_2 \mu - \alpha) \\
&\le \frac{1}{(\mu (2^\alpha-1))^{2m}} \left(4 \left(\frac{4 m}{C_1 d}\right)^m + \frac{\eps}{d^k}(d^4 + \mu)^{2m} \right).
\eas
We now simplify the right hand side.  Let $\delta = 2^\alpha - 1$.  For $d_S \ge 2$, we have $\mu \ge 1/d_S$.  We shall also assume that $m = k/8$.  This gives us (using $\mu \le 1$)
\be
\Pr_\nu(S(\psi_S) \le \log_2 d_S - \alpha - \beta) \le \left(\frac{d_S}{\delta}\right)^{k/4} \left(4 \left(\frac{k}{2C_1 d}\right)^{k/8} + \eps\left(1 + \frac{1}{d^4}\right)^{k/4} \right).
\ee
Now, one can easily show (e.g.~by induction on $n$) that
\be
(1+\delta)^n \le 2
\ee
for $2n\delta \le 1$.  We use this for $n = k/4$ and $\delta = 1/d^4$.  The condition is then $k \le 2 d^4$, which we shall assume (we will set $k = \log d/\log \log d$ later).  We now obtain
\be
\Pr_\nu(S(\psi_S) \le \log_2 d_S - \alpha - \beta) \le \left(\frac{d_S}{\delta}\right)^{k/4} \left(4 \left(\frac{k}{2C_1 d}\right)^{k/8} + 2\eps \right).
\ee
We will now take $\eps = 2 \left(\frac{k}{2C_1 d}\right)^{k/8}$, so that the two terms are the same.  $\log 1/\eps$ is $\poly \log d$ so this remains efficient.  Now
\be
\Pr_\nu(S(\psi_S) \le \log_2 d_S - \alpha - \beta) \le 8 \left(\frac{d_S^2 k}{2 C_1 d \delta^2}\right)^{k/8}.
\ee
Assuming that $\delta^2 > \frac{k d_S^2}{2 C_1 d}$, we should take $k$ as large as possible up to $\frac{2 C_1 \delta^2 d}{e d_S^2}$, when the right hand side is maximised.  We then find the result after further simplification.
\end{proof}


\begin{thebibliography}{10}

\bibitem{AlonNussboim08}
N.~{Alon} and A.~{Nussboim}.
\newblock k-wise independent random graphs.
\newblock {\em 49th Annual IEEE Symposium on Foundations of Computer Science},
  0:813--822, 2008.
\newblock arXiv:0804.1268.

\bibitem{AmbainisEmerson07}
A.~{Ambainis} and E.~{Emerson}.
\newblock {Quantum t-designs: t-wise independence in the quantum world}.
\newblock {\em IEEE Conference on Computational Complexity 2007}, 2007.
\newblock arXiv:quant-ph/0701126v2.

\bibitem{AubrunRandomizingChannels}
G.~{Aubrun}.
\newblock {On almost randomizing channels with a short Kraus decomposition},
  2008.
\newblock arXiv:0805.2900.

\bibitem{BarnumLinden01}
H.~{Barnum} and N.~{Linden}.
\newblock Monotones and invariants for multi-particle quantum states.
\newblock {\em Journal of Physics A}, 34(35):6787--6805, 2001.

\bibitem{BellareRompel}
M.~{Bellare} and J.~{Rompel}.
\newblock Randomness-efficient oblivious sampling.
\newblock {\em 35th Annual IEEE Symposium on Foundations of Computer Science},
  pages 276--287, Nov 1994.

\bibitem{MostStatesUselessMBQCBMW}
M.~J. {Bremner}, C.~{Mora}, and A.~{Winter}.
\newblock {Are random pure states useful for quantum computation?}, 2008.
\newblock arXiv:0812.3001.

\bibitem{DahlstenPlenio05}
O.~{Dahlsten} and M.~{Plenio}.
\newblock Entanglement probability distribution of bipartite randomised
  stabilizer states.
\newblock {\em Quant. Inf. Comp.}, 6(6):527--538, 2006.
\newblock arXiv:quant-ph/0511119.

\bibitem{DCEL06}
C.~{Dankert}, R.~{Cleve}, J.~{Emerson}, and E.~{Livine}.
\newblock {Exact and Approximate Unitary 2-Designs: Constructions and
  Applications}, 2006.
\newblock arXiv:quant-ph/0606161.

\bibitem{Locking04}
D.~P. {DiVincenzo}, M.~{Horodecki}, D.~W. {Leung}, J.~A. {Smolin}, and B.~M.
  {Terhal}.
\newblock Locking classical correlations in quantum states.
\newblock {\em Phys. Rev. Lett.}, 92(6):067902, Feb 2004.
\newblock arXiv:quant-ph/0303088.

\bibitem{PagesConjectureProof94}
S.~K. {Foong} and S.~{Kanno}.
\newblock Proof of page's conjecture on the average entropy of a subsystem.
\newblock {\em Phys. Rev. Lett.}, pages 1148--1151, 1994.

\bibitem{MostStatesUselessMBQCGFE}
D.~{Gross}, S.~{Flammia}, and J.~{Eisert}.
\newblock {Most quantum states are too entangled to be useful as computational
  resources}, 2008.
\newblock arXiv:0810.4331.

\bibitem{SuperdenseCodingHHL}
A.~{Harrow}, P.~{Hayden}, and D.~{Leung}.
\newblock Superdense coding of quantum states.
\newblock {\em Phys. Rev. Lett.}, 92(18):187901, May 2004.
\newblock arXiv:quant-ph/0307221.

\bibitem{EfficientTPE08}
A.~W. {Harrow} and R.~A. {Low}.
\newblock Efficient quantum tensor product expanders and $k$-designs, 2008.
\newblock arXiv:0811.2597.

\bibitem{RandomCircuits08}
A.~W. {Harrow} and R.~A. {Low}.
\newblock Random quantum circuits are approximate 2-designs, 2008.
\newblock arXiv:0802.1919.

\bibitem{RandomizingQuantumStates04}
P.~{Hayden}, D.~{Leung}, P.~W. {Shor}, and A.~{Winter}.
\newblock {Randomizing Quantum States: Constructions and Applications}.
\newblock {\em Communications in Mathematical Physics}, 250:371--391, 2004.
\newblock arXiv:quant-ph/0307104.

\bibitem{AspectsOfGenericEntanglement}
P.~{Hayden}, D.~W. {Leung}, and A.~{Winter}.
\newblock {Aspects of Generic Entanglement}.
\newblock {\em Communications in Mathematical Physics}, 265:95--117, July 2006.
\newblock arXiv:quant-ph/0407049.

\bibitem{Ledoux}
M.~{Ledoux}.
\newblock {\em {The Concentration of Measure Phenomenon}}.
\newblock {American Mathematical Society}, 2001.

\bibitem{PagesConjecture}
D.~N. {Page}.
\newblock Average entropy of a subsystem.
\newblock {\em Phys. Rev. Lett.}, 71:1291, 1993.

\bibitem{ThermalisationPSW}
S.~{Popescu}, A.~J. {Short}, and A.~{Winter}.
\newblock Entanglement and the foundations of statistical mechanics.
\newblock {\em Nature Physics}, 2:754--758, 2006.
\newblock arXiv:quant-ph/0511225.

\bibitem{PagesConjectureProof95}
J.~{Sanchez-Ruiz}.
\newblock {Simple proof of Page's conjecture on the average entropy of a
  subsystem}.
\newblock {\em Phys. Rev. E}, page 5653, 1995.

\bibitem{GeometricEntanglementShimony}
A.~{Shimony}.
\newblock {Degree of Entanglement}.
\newblock {\em Ann. N.Y. Acad. Sci.}, 755:675, 1995.

\bibitem{SmithLeung06}
G.~{Smith} and D.~{Leung}.
\newblock Typical entanglement of stabilizer states.
\newblock {\em Phys. Rev. A}, 74(6):062314, 2006.
\newblock arXiv:quant-ph/0510232.

\end{thebibliography}
\end{document}